# NaSrMn$_2$F$_7$, NaCaFe$_2$F$_7$, and NaSrFe$_2$F$_7$: novel single crystal pyrochlore antiferromagnets

M.B. Sanders*, J.W. Krizan*, K.W. Plumb, T.M. McQueen, and R.J. Cava
*Department of Chemistry, Princeton University, Princeton, New Jersey 08544*
*these authors contributed equally to this work

**Abstract**

The crystal structures and magnetic properties of three previously unreported A$_2$B$_2$F$_7$ pyrochlore materials, NaSrMn$_2$F$_7$, NaCaFe$_2$F$_7$, and NaSrFe$_2$F$_7$ are presented. In these compounds, either S=2 Fe$^{2+}$ or S=5/2 Mn$^{2+}$ is on the B site, while nonmagnetic Na and Ca (Na and Sr) are disordered on the A site. The materials, which were grown as crystals via the floating zone method, display high effective magnetic moments and large Curie-Weiss thetas. Despite these characteristics, freezing of the magnetic spins, characterized by peaks in the susceptibility or specific heat, is not observed until low temperatures. The empirical frustration index, f=- θcw/T$_f$, for the materials are 36 (NaSrMn$_2$F$_7$), 27 (NaSrFe$_2$F$_7$), and 19 (NaCaFe$_2$F$_7$). AC susceptibility, DC susceptibility, and heat capacity measurements are used to characterize the observed spin glass behavior. The results suggest that the compounds are frustrated pyrochlore antiferromagnets with weak bond disorder. The magnetic phenomena that these fluoride pyrochlores exhibit, in addition to their availability as relatively large single crystals, make them promising candidates for the study of geometric magnetic frustration.

**Introduction**

The 3D pyrochlore is a model system for studying the consequences of geometric magnetic frustration. Indeed, the A and B ions in the $A_2B_2X_7$ pyrochlore structure each take on their own lattice of corner-sharing tetrahedra; systems with magnetic ions on either the A or B site may exhibit magnetic frustration. Rare earth pyrochlore oxides have been of recent interest, often with Ti or Sn on the nonmagnetic B site (i.e., materials of the type $RE_2(Ti/Sn)_2O_7$). These compounds display a variety of interesting magnetic phenomena, such as spin ice ($Dy_2Ti_2O_7$, $Ho_2Sn_2O_7$, $Ho_2Ti_2O_7$), spin liquid ($Tb_2Ti_2O_7$), long range ordered ($Gd_2Ti_2O_7$, $GdSn_2O_7$, $Er_2Ti_2O_7$), and spin glass ($Y_2Mo_2O_7$, $Tb_2Mo_2O_7$) behavior.[1,2,3,4,5]

Rare earth pyrochlores are well-known for their low magnetic coupling strengths, which add a layer of complexity to the interpretation of their magnetic properties, as small changes in structure or the influence of what are nominally weak auxiliary interactions can significantly affect their ground state at low temperatures. Investigation of pyrochlore systems containing elements with larger magnetic interactions, such as transition metals, may therefore present alternative pathways in the study of magnetic frustration.[6] This has motivated us to examine fluoride pyrochlores of the $(A^+A'^{2+})B_2^{2+}F_7$ type, where B is a transition metal, where the mixture of A and A' atoms with different charges is required to preserve charge neutrality. Several of these fluorine-based pyrochlores have been successfully grown as single crystals and characterized, and yet there exists only a limited amount of information on the properties of materials in this family.[7,8,9] Here we report the crystal structures and magnetic properties of three previously unreported fluoride pyrochlore materials: $NaSrMn_2F_7$, $NaCaFe_2F_7$, and $NaSrFe_2F_7$. The three pyrochlores display a high effective moment for Fe or Mn (5.94 $\mu_B$/Fe for $NaSrFe_2F_7$; 5.61 $\mu_B$/Fe for $NaCaFe_2F_7$; and 6.25 $\mu_B$/Mn for $NaSrMn_2F_7$), a high Weiss Temperature (-98.1 K, $NaSrFe_2F_7$; -72.8 K, $NaCaFe_2F_7$; -89.7 K, $NaSrMn_2F_7$), and no spin freezing until roughly ~3-4 K for the iron pyrochlores and ~2-3 K for the manganese compound, indicating that these materials are highly magnetically frustrated.

Structurally, Na and Ca (Na and Sr) are fully disordered on the A site of the pyrochlore with the Fe and F (Mn and F) sublattices fully ordered. The magnetic properties suggest that the A-site disorder may produce weak bond disorder in the Fe-Fe (Mn-Mn) exchange interactions. Such bond disorder has been analyzed from a theoretical approach, and has been proposed to explain the properties of the spin glass $Y_2Mo_2O_7$.[10] However, the chemical cause of the bond disorder in the fluoride systems investigated here—Na/Ca or Na/Sr mixing on the B site—is much more apparent.

Like the previously reported $NaCaNi_2F_7$, $NaCaCo_2F_7$, and $NaSrCo_2F_7$ fluoride pyrochlores, $NaSrMn_2F_7$, $NaCaFe_2F_7$, and $NaSrFe_2F_7$ were grown as single crystals by the floating zone method.[7,8,9] The successful production of these compounds suggests that transition metal fluorides have the potential to develop into a large new family of pyrochlore materials to advance the study of geometric frustration.

**Experimental**

Single crystals of $NaSrMn_2F_7$, $NaCaFe_2F_7$, and $NaSrFe_2F_7$ were prepared in an optical floating zone furnace. Single crystal x-ray diffraction determination of the crystal structures at 293 K were performed on a Bruker diffractometer equipped with an Apex II detector and graphite-monochromated Mo Kα radiation. The crystals were crushed into fine powders and then analyzed by powder x-ray diffraction on a Bruker D8 Advance Eco diffractometer with Cu Kα radiation (λ=1.5418 Å) and a Lynxeye detector. Temperature and field dependent magnetization measurements of the powders were made using the ACMS option in a Quantum Design Physical Property Measurement System (PPMS) and Quantum Design superconducting quantum interference device (SQUID). Heat capacity measurements were carried out using the heat relaxation method in the PPMS; 1-3 mg single crystal samples were mounted on a nonmagnetic sapphire stage with Apiezon N grease. The nonmagnetic analog $NaCaZn_2F_7$ was used for the subtraction to estimate the magnetic contribution to the specific heat in $NaSrMn_2F_7$ and $NaSrFe_2F_7$.

## Results

### Structure Solutions

Tables 1-4 summarize the single-crystal structure solutions for of $NaSrFe_2F_7$, $NaCaFe_2F_7$, and $NaSrMn_2F_7$. All three compounds are previously unreported, fully fluorinated, stoichiometric pyrochlores of the $A_2B_2F_7$ type where Na and Sr or Na and Ca are completely disordered on the A site and $Fe^{2+}$ or $Mn^{2+}$ completely occupies the B site. The reciprocal lattice images in Figure 1 for each fluoride show no indication of any long range or short range Na/Sr or Na/Ca ordering on the B site. Similarly, the single crystal solution did not detect any A-B site mixing. As observed in many other pyrochlores, the B-site octahedron ($FeF_6$ or $MnF_6$) is slightly compressed along the <111> direction and has six equal bond lengths (Figure 2).[7,8,9] This compression results in two F-Mn-F or F-Fe-F bond angles that deviate from the ideal 90° (Table 5). Of the three pyrochlores under investigation here, the angles for $NaSrFe_2F_7$ deviate the least, which suggests that the local crystal field around the transition metal is the most regular. Figure 1 also depicts powder x-ray diffraction patterns from ground single crystals, all of which show a characteristic pyrochlore pattern. To the top right of each pattern is an image of the single crystals that were employed in this study.

### Magnetic Measurements

The temperature dependent bulk DC susceptibility of $NaSrFe_2F_7$, $NaCaFe_2F_7$, and $NaSrMn_2F_7$ are shown in Figures 3-5, respectively. The magnetic data for the compounds were fit to the Curie-

Weiss law $\chi = \frac{C}{T-\theta_{CW}}$ where $\chi$ is the magnetic susceptibility, C is the Curie Constant, and $\theta_{CW}$ is the Weiss temperature. The effective moments were then obtained using $\mu_{eff} \propto 2.83\sqrt{C}$. The frustration index $f = \frac{-\theta_{CW}}{T_C}$ was used compare the strength of magnetic frustration among all three compounds, where $T_c$ is the transition temperature. The applied field (H)-dependent magnetization (M) plots of each compound at 2 K are shown in the upper insets in Figures 3-5. For the three samples under investigation, the M vs. H behavior does not deviate significantly from a straight line near 2000 Oe, and so $\chi$ is defined as M/H where H= 2000 Oe in all susceptibility measurements. The results of the susceptibility measurements are summarized in Table 6.

**NaSrFe$_2$F$_7$ DC Susceptibility**

The magnetic susceptibility and inverse susceptibility for NaSrFe$_2$F$_7$ are shown in Figure 3. The field-dependent magnetization *M(H)* at 2 K (upper insert) shows subtle curvature up to 9 T, but the magnetization is nowhere near saturation within this range. Data were fit to the Curie-Weiss law (main panel) in the temperature range 200-300 K, yielding a Weiss temperature of -98.1 K and an effective moment of 5.94 $\mu_B$/Fe. The moment is larger than the spin only value (4.90 $\mu_B$) for S=2 Fe$^{2+}$, indicating the presence of a significant orbital contribution. The negative Weiss temperature is suggestive of strong antiferromagnetic interactions, though no features are seen in the susceptibility until 3.7 K. This gives a frustration index *f* of 26.5.

**NaCaFe$_2$F$_7$ DC Susceptibility**

The $\chi(T)$ plot of NaCaFe$_2$F$_7$ is shown in the main panel of Figure 4. Fitting the data to the Curie-Weiss law in the 200-300 K temperature range gives a Weiss temperature of -72.8 K and a moment of 5.61 $\mu_B$/Fe. Similar to that observed for NaSrFe$_2$F$_7$, the moment for NaCaFe$_2$F$_7$ is higher than the spin-only value of 4.90 $\mu_B$ for S=2 Fe$^{2+}$. However, it is close to the observed moment, which typically ranges from 5.0 to 5.6 $\mu_B$. Once again, the large magnitude of the negative Weiss temperature is indicative of antiferromagnetic exchange interactions. The frustration index was calculated using the $\theta_{CW}$ obtained from the Curie-Weiss fit and the feature observed at 3.9 K in the magnetic susceptibility, yielding a value of 18.7. The applied field (H)-dependent magnetization (M) at 2 K (upper inset) shows slight curvature but no signs of saturation, nearly identical to that seen for $NaSrFe_2F_7$.

**$NaSrMn_2F_7$ DC Susceptibility**

In the top inset of Figure 5 is the 2K field-dependent magnetization $M(H)$ of $NaSrMn_2F_7$. The response is linear and reversible with the field increasing and decreasing and shows no signs of saturation. The main panel of Figure 5 displays the magnetic susceptibility of $NaSrMn_2F_7$ in an applied field of 2000 Oe and in the lower inset the inverse susceptibility. Data were fit to the Curie-Weiss law from 200-300 K, resulting in a Weiss Temperature of -89.7 K and an effective moment of 6.25 $\mu_B$/Mn. The effective moment per manganese is larger than the spin only moment (5.92 $\mu_B$) for S=5/2 $Mn^{2+}$, which may be attributed to experimental error. The large negative $\theta_{CW}$ is indicative of strong antiferromagnetic interactions between $Mn^{2+}$ spins, and no magnetic transition is apparent in the susceptibility until approximately 2.5 K. This presents a frustration index $f$ of 35.9.

The Curie-Weiss law was rearranged $\frac{C}{\chi|\theta|} = \frac{T}{|\theta|} - 1$ in Figure 6, enabling the formation of a normalized, dimensionless plot, useful for comparing the magnetic behavior of the fluoride pyrochlores. Ideal antiferromagnets would follow the black dashed line, y=x+1, where magnetic ordering is on the order of $T/|\theta| \approx 1$ (Figure 6). Using this as a model, the magnetic behavior of $NaCaFe_2F_7$, $NaSrFe_2F_7$, and $NaSrMn_2F_7$ is compared with three other fluoride pyrochlores: $NaSrCo_2F_7$, $NaCaNi_2F_7$, and $NaCaCo_2F_7$. While it is evident that all six compounds follow ideal Curie-Weiss behavior at high temperatures (the dashed line), they each deviate at low temperatures. This is highlighted in the inset to Figure 6. Antiferromagnetic or ferromagnetic correlations at lower temperatures are displayed as positive and negative deviations from the y=x+1 Curie-Weiss model, respectively. Both NaSrFe$_2$F$_7$ and NaCaFe$_2$F$_7$ deviate from ideal Curie-Weiss behavior with large ferromagnetic fluctuations that begin at roughly 1.50 T/|$\theta$| (NaSrFe$_2$F$_7$) and 0.75 T/|$\theta$| (NaCaFe$_2$F$_7$) and become stronger as they approach their spin glass transition. NaSrMn$_2$F$_7$ also experiences slight ferromagnetic fluctuations, from about 1.20 to 0.65 T/|$\theta$| that increase in intensity before its spin freezing transition. While NaSrCo$_2$F$_7$'s ferromagnetic divergence from Curie-Weiss behavior around its spin glass ordering is similar to that of NaSrMn$_2$F$_7$, the cobalt analog experiences antiferromagnetic fluctuations at slightly higher temperatures that start at ~0.08 T/|$\theta$|. NaCaCo$_2$F$_7$ also experiences ferromagnetic deviations before spin freezing, but not to the same extent that the other fluorides do. NaCaNi$_2$F$_7$, conversely, exhibits antiferromagnetic fluctuations at low temperatures as it approaches its spin ordering transition. For all six fluoride pyrochlores in Figure 6, the deviations they exhibit from ideal Curie-Weiss behavior above their spin freezing transitions suggest the presence of competing interactions within their magnetic sublattice.

**AC Susceptibility**

Figure 7 further examines the low temperature magnetization of NaCaFe$_2$F$_7$, NaSrFe$_2$F$_7$, and NaSrMn$_2$F$_7$ near the magnetic freezing, which occurs at ~3.7-3.9 K for the iron pyrochlores and ~2.5 K for the manganese pyrochlore. The top two panels show the temperature dependent ac susceptibility at a number of different frequencies from 3-5.5 K and in the lower left panel from 1-3.4 K. No features are observed in the AC susceptibility beyond the temperature range selected for each compound. A systematic upward temperature shift in the spin freezing is observed with increasing frequency in each plot in Figure 7. This is strong evidence for a glass-like phase transition in the magnetic system. The expression $\frac{\Delta T_f}{T_f \Delta log\omega}$ describes the shift of the peak temperature as a function of frequency, and can be used to characterize spin-glass and spin-glass-like materials.[11] These results are similar to those of several other reported fluoride pyrochlores (0.027, $NaSrCo_2F_7$; 0.029, $NaCaCo_2F_7$; 0.024, $NaCaNi_2F_7$) and fall within the range expected for an insulating spin glass.[7,8,9,12,13]

The spin glass behavior can be further characterized by fitting the shift in the AC susceptibility to the Vogel-Fulcher law, $T_f = T_0 - \frac{E_a 1}{k_b \ln(t_0 f)}$, which relates the freezing temperature and frequency to the intrinsic relaxation time ($t_0$), the activation energy of the process ($E_a$), and the ideal glass temperature ($T_0$). The fits for $NaCaFe_2F_7$, $NaSrFe_2F_7$, and $NaSrMn_2F_7$ are compared in the lower right inset of Figure 7 and the results are summarized in Table 7. In order to compare the materials, $T_f$ is scaled on the y-axis by the freezing temperature of each compound. Due to the limited temperature resolution and few data points available, a meaningful fit to the parameters could not be obtained for the compounds under investigation. Therefore, $t_0$ was set to a physically meaningful value of $1 \times 10^{-12}$ s in each calculation, allowing for direct comparison among the three compounds. For $NaCaFe_2F_7$, values of $2.9(4) \times 10^{-3}$ eV and $2.8(4)$ K were obtained for $E_a$ and $T_0$, respectively. The fit for $NaSrFe_2F_7$ yielded $E_a = 2.8(1) \times 10^{-3}$ eV and $T_0 = 2.7(2)$ K. For $NaSrMn_2F_7$, an activation energy of $8.3(4) \times 10^{-4}$ ev and an ideal glass temperature of $2.2(7)$ K were obtained. In each case, the maximum in the magnetic transition observed in the DC susceptibility of each compound is estimated to be the freezing temperature and corresponds well with the calculated value of $T_0$. The activation energies are close to those of other reported fluoride pyrochlores (i.e., $1.4(2) \times 10^{-3}$ eV for $NaCaNi_2F_7$; $1.3(1) \times 10$-3 eV for $NaSrCo_2F_7$; and $1.0 \times 10$-3 eV for $NaCaCo_2F_7$).[8,9,7] A comparison of the fits in Figure 7 reveals the similarity among the three materials. A slightly higher activation energy was obtained for $NaCaFe_2F_7$ than $NaSrFe_2F_7$; yet, on a relative scale of $T_f/T_N$, $NaCaFe_2F_7$'s ideal glass temperature is lower. More detailed measurements are necessary to obtain a better understanding of the differences inherent in these compounds.

**Heat Capacity**

The magnetic spin behavior was further analyzed through heat capacity measurements. The magnetic heat capacity of NaSrMn$_2$F$_7$ and NaSrFe$_2$F$_7$ (Figure 8) were obtained by subtracting the heat capacity of the nonmagnetic analog, NaCaZn$_2$F$_7$. The heat capacity of NaCaZn$_2$F$_7$ was scaled by 20% as an estimate of the phonon contribution. From roughly 20 to 2 K, NaSrMn$_2$F$_7$'s magnetic heat capacity is much greater, approaching the spin freezing temperature. Below 5 K for both compounds, there is a broad maximum in the magnetic heat capacity slightly lower in temperature than the transition observed in each compound's AC susceptibility. This peak shape and position are similar to that observed for NaCaCo$_2$F$_7$, NaCaNi$_2$F$_7$, and NaSrCo$_2$F$_7$, which is indicative of spin glass behavior.[7,8,9,11] Both compounds show a saturation of the magnetic heat capacity by roughly 30 K.

By probing the low temperature heat capacity below the transition, it is possible to gain information about the elementary magnetic excitations and infer the ground state.[14] Spin glasses typically exhibit a linear heat capacity on a $C(T) \propto T$ scale, resulting from a distribution of two level systems linear in energy.[15,14] Some curvature is apparent for both NaSrFe$_2$F$_7$ and NaSrMn$_2$F$_7$ in the top panel of Figure 8 plotted as C$_{mag}$/T versus T; they are also nonlinear on T$^2$ or T$^{3/2}$ scales. Therefore, the current data does not allow for any conclusions to be made regarding the low temperature spin excitations.

Integration of the magnetic heat capacity allows for an estimation of the magnetic entropy freezing out in the transition (lower panel of Figure 8). The value for NaSrFe$_2$F$_7$ is very close to that of a two-state magnetic system, R ln(2S+1); however, NaSrMn$_2$F$_7$ releases more entropy than can be accounted for by its Heisenberg limit, Rln(6). While its proximity to R ln(2S+1) suggests that it may be a 2-state system, additional work is necessary to determine this conclusively.

**Conclusion**

While the large negative Weiss temperatures obtained from the Curie Weiss fits suggest strong antiferromagnetic interactions, no spin freezing is observed until very low temperatures for $NaCaFe_2F_7$, $NaSrFe_2F_7$, and $NaSrMn_2F_7$. This indicates that the materials are highly frustrated. Fitting the AC susceptibility data near the spin freezing temperature with the Vogel-Fulcher law yields results that are consistent with common spin glass materials. The spin glass behavior is verified by heat capacity measurements, which indicate that $NaSrFe_2F_7$ has a total entropy loss near R ln(4), arising from spin freezing. Additional work is necessary to further characterize the heat capacity of $NaSrMn_2F_7$, and future heat capacity measurements need to be conducted on $NaCaFe_2F_7$. The crystal structure determinations indicate that Na and Ca (or Na and Sr) are disordered on the A-site. As has been observed for other pyrochlores, disorder produces a random local environment around the B-site cation.[16] Theoretical calculations of the Heisenberg antiferromagnet on the pyrochlore have shown that weak exchange randomness (disorder) can instigate spin freezing with the disorder strength proportional to the spin glass temperature.[17,18] This is likely the case for the pyrochlores involved in this study. Further analysis of these materials' magnetic properties may be of significant interest; the availability of large single crystals will help advance future investigations.

**Acknowledgements**

All of this research was supported by the U.S. Department of Energy, Division of Basic Energy Sciences, Grant No. DE-FG02-08ER46544, through the Institute for Quantum Matter at Johns Hopkins University. The authors are thankful for discussions with Michel Gingras, Leon Balents, Lucile Savary and Roderich Moessner that inspired this work and our previous work on transition metal fluoride pyrochlores.

**Figure Captions**

**Figure 1:** Powder diffraction patterns of crushed $NaSrMn_2F_7$, $NaCaFe_2F_7$, and $NaSrFe_2F_7$ single crystals, confirming that the bulk material is an $A_2B_2F_7$ pyrochlore. On the left, above each powder pattern, is a precession image from a single crystal diffraction of the (0kl) plane. No long range structural ordering, which would be seen as superlattice spots inconsistent with the *Fd-3m* pyrochlore space group, is evident in these images. On the right side of the figure are slices of the single crystals that were employed in the study.

**Figure 2:** An $FeF_6$ octahedron in $NaSrFe_2F_7$ displaying compression along the <111> axis. Although all six Fe-F bond lengths are equivalent (2.07 Å), there are two different bond angles: 97.1° and 82.9°.

**Figure 3:** In the main panel is the inverse DC susceptibility of $NaSrFe_2F_7$ in an applied field of 2000 Oe. The Curie-Weiss fit is shown in black. The lower right inset shows the raw susceptibility. The upper left inset displays the magnetization as a function of applied field at 2 K.

**Figure 4:** The inverse DC susceptibility of $NaCaFe_2F_7$ in an applied field of 2000 Oe (main panel). The Curie-Weiss fit is shown in yellow. The raw DC susceptibility is displayed in the lower right inset. In the upper left inset is the magnetization at 2 K as a function of applied field.

**Figure 5:** Main panel: the temperature-dependent inverse DC susceptibility of $NaSrMn_2F_7$ in an applied field of 2000 Oe. The Curie-Weiss fit is shown in light pink. Lower right panel: the raw DC susceptibility in an applied field of 2000 Oe. Upper left panel: the magnetization as a function of applied field at 2 K.

**Figure 6:** Main Panel: A dimensionless, normalized plot of the magnetization, comparing $NaSrMn_2F_7$, $NaCaFe_2F_7$, and $NaSrFe_2F_7$ with three additional fluoride pyrochlores ($NaSrCo_2F_7$, $NaCaNi_2F_7$, and $NaCaCo_2F_7$) at temperatures near and below the Weiss temperature. This was achieved by rearranging the Curie-Weiss law like so: $\frac{C}{\chi|\theta|} = \frac{T}{|\theta|} - 1$. The fluorides deviate from Curie-Weiss behavior at low

temperatures, as indicated by the black dashed line. The lower right inset highlights the low T/$|\theta|$ near the transitions.

**Figure 7:** The frequency dependence of the AC susceptibility in an applied field of 20 Oe for NaCaFe$_2$F$_7$ (upper left), NaSrFe$_2$F$_7$ (upper right), and NaSrMn$_2$F$_7$ (lower left). The bottom right panel shows the parameterization of the shift of the AC susceptibility by fitting the frequency-dependent freeing temperature to the Vogel-Fulcher law. The fit is displayed for all three pyrochlores for comparison.

**Figure 8:** The heat capacity per mole of the B-site cation (Fe or Mn). The inset to the top panel shows the heat of capacity of NaSrMn$_2$F$_7$ and NaSrFe$_2$F$_7$. The heat capacity of the nonmagnetic analog NaCaZn$_2$F$_7$ was subtracted from that of NaSrMn$_2$F$_7$ and NaSrFe$_2$F$_7$ to obtain the magnetic heat capacity. The top main panel shows the temperature dependence of the heat capacity below the spin-freezing transition. In the bottom panel, the entropy from the magnetic heat capacity is compared to the Ising (R ln(2)) and Heisenberg (R (ln(2S+1))) limits.

## Tables
Table 1: Single Crystal Data and Structural Refinement for $NaSrMn_2F_7$, $NaSrFe_2F_7$, and $NaCaFe_2F_7$

|  | $NaSrMn_2F_7$ | $NaSrFe_2F_7$ | $NaCaFe_2F_7$ |
|---|---|---|---|
| **Formula Weight (g/mol)** | 353.49 | 355.31 | 307.77 |
| **Crystal System** | Cubic | Cubic | Cubic |
| **Space Group** | *Fd-3m* | *Fd-3m* | *Fd-3m* |
| **Unit Cell (Å)** | 10.8012(3) | 10.6323(4) | 10.4471(2) |
| **Volume (Å$^3$)** | 1260.13(10) | 1201.94(14) | 1140.22(7) |
| **Z** | 8 | 8 | 8 |
| **Radiation** | MoKα | MoKα | MoKα |
| **Temperature (K)** | 293(2) | 293(2) | 293(2) |
| **Adsorption Coefficient** | 12.517 | 13.749 | 6.137 |
| **F(000)** | 1296 | 1312 | 1168 |
| **Reflections Collected/ Unique/internal data agreement** | 6248/163 $R_{int}$=0.0234 | 5699/154 $R_{int}$=0.0316 | 4887/147 $R_{int}$=0.0331 |
| **Data/Parameters** | 163/11 | 154/13 | 147/10 |
| **Goodness-of-fit** | 1.223 | 1.083 | 1.216 |
| **Final R Indiced [l>2σ(l)]** | $R_1$= 0.0259 $wR_2$=0.0499 | R1= 0.0286 $wR_2$=0.1007 | R1=0.0313 $wR_2$=0.0466 |
| **Largest diff. peak and hole** | 0.585 and -0.543 e Å$^{-3}$ | 0.890 and -0.683 e Å$^{-3}$ | 0.391 and -0.341 e Å$^{-3}$ |

**Table 2: Atomic positions and anisotropic thermal displacement parameters for NaSrMn$_2$F$_7$**

| | Site | x | y | z | Occ. | U11 | U22 | U33 | U23 | U13 | U12 |
|---|---|---|---|---|---|---|---|---|---|---|---|
| Na | 16d | 0.5 | 0.5 | 0.5 | 0.5 | 0.0216(3) | 0.0216(3) | 0.0216(3) | -0.00358(18) | -0.00358(18) | -0.00358(18) |
| Sr | 16d | 0.5 | 0.5 | 0.5 | 0.5 | 0.0216(3) | 0.0216(3) | 0.0216(3) | -0.00358(18) | -0.00358(18) | -0.00358(18) |
| Mn | 16c | 0 | 0 | 0 | 1 | 0.0106(2) | 0.0106(2) | 0.0106(2) | -0.00008(13) | -0.00008(13) | -0.00008(13) |
| F(1) | 8b | 0.375 | 0.375 | 0.375 | 1 | 0.0233(11) | 0.0233(11) | 0.0233(11) | 0 | 0 | 0 |
| F(2) | 48f | 0.3331(2) | 0.125 | 0.125 | 1 | 0.0369(13) | 0.0210(6) | 0.0210(6) | 0.0104(8) | 0 | 0 |

**Table 3: Atomic positions and anisotropic thermal displacement parameters for NaSrFe$_2$F$_7$**

| | Site | x | y | z | Occ. | U11 | U22 | U33 | U23 | U13 | U12 |
|---|---|---|---|---|---|---|---|---|---|---|---|
| Na | 16d | 0.5 | 0.5 | 0.5 | 0.5 | 0.081(6) | 0.081(6) | 0.081(6) | -0.012(3) | -0.012(3) | -0.012(3) |
| Sr | 16d | 0.5 | 0.5 | 0.5 | 0.5 | 0.0154(5) | 0.0154(5) | 0.0154(5) | -0.0024(2) | -0.0024(2) | -0.0024(2) |
| Fe | 16c | 0 | 0 | 0 | 1 | 0.0085(3) | 0.0085(3) | 0.0085(3) | -0.00027(9) | -0.00027(9) | -0.00027(9) |
| F(1) | 8b | 0.375 | 0.375 | 0.375 | 1 | 0.0230(9) | 0.0230(9) | 0.0230(9) | 0 | 0 | 0 |
| F(2) | 48f | 0.3313(2) | 0.125 | 0.125 | 1 | 0.0244(11) | 0.0169(6) | 0.0169(6) | 0.0071(6) | 0 | 0 |

**Table 4: Atomic positions and anisotropic thermal displacement parameters for NaCaFe$_2$F$_7$**

| | Site | x | y | z | Occ. | U11 | U22 | U33 | U23 | U13 | U12 |
|---|---|---|---|---|---|---|---|---|---|---|---|
| Na | 16d | 0.5 | 0.5 | 0.5 | 0.5 | 0.0166(3) | 0.0166(3) | 0.0166(3) | -0.0027(3) | -0.0027(3) | -0.0027(3) |
| Ca | 16d | 0.5 | 0.5 | 0.5 | 0.5 | 0.0166(3) | 0.0166(3) | 0.0166(3) | -0.0027(3) | -0.0027(3) | -0.0027(3) |
| Fe | 16c | 0 | 0 | 0 | 1 | 0.00866(17) | 0.00866(17) | 0.00866(17) | -0.00005(14) | -0.00005(14) | -0.00005(14) |
| F(1) | 8b | 0.375 | 0.375 | 0.375 | 1 | 0.0180(9) | 0.0180(9) | 0.0180(9) | 0 | 0 | 0 |
| F(2) | 48f | 0.33702(18) | 0.125 | 0.125 | 1 | 0.0238(9) | 0.0184(5) | 0.0184(5) | 0.0085(7) | 0 | 0 |

**Table 5: Bond Distances and Angles**

| | NaSrMn$_2$F$_7$ | NaSrFe$_2$F$_7$ | NaCaFe$_2$F$_7$ |
|---|---|---|---|
| (Mn/Fe)-F (Å) | 2.11 | 2.07 | 2.06 |
| F-(Mn/Fe)-F (°) | 97.8 and 82.2 | 97.1 and 82.9 | 99.1 and 80.9 |

**Table 6: Data from Curie-Weiss fits: temperature range for fit; Weiss temperature; magnetic moment**

| Compound | Temperature Range/ K | Weiss Temperature/ K | Magnetic Moment/ $\mu_B$ |
|---|---|---|---|
| $NaSrMn_2F_7$ | 200-300 | -89.72 | 6.25 |
| $NaSrFe_2F_7$ | 200-300 | -98.12 | 5.94 |
| $NaCaFe_2F_7$ | 200-300 | -72.79 | 5.61 |

**Table 7: Values from Vogel-Fulcher fits: intrinsic relaxation time ($t_0$), the activation energy ($E_a$), and ideal glass temperature ($T_0$)**

| Compound | $t_0$ (s) | $E_a$ (ev) | $T_0$ (K) |
|---|---|---|---|
| $NaSrMn_2F_7$ | $1\times10^{-12}$ | $8.3(4)\times10^{-4}$ | 2.2(7) |
| $NaSrFe_2F_7$ | $1\times10^{-12}$ | $2.8(1)\times10^{-3}$ | 2.7(2) |
| $NaCaFe_2F_7$ | $1\times10^{-12}$ | $2.9(4)\times10^{-3}$ | 2.8(4) |

**Figures**
**Figure 1**

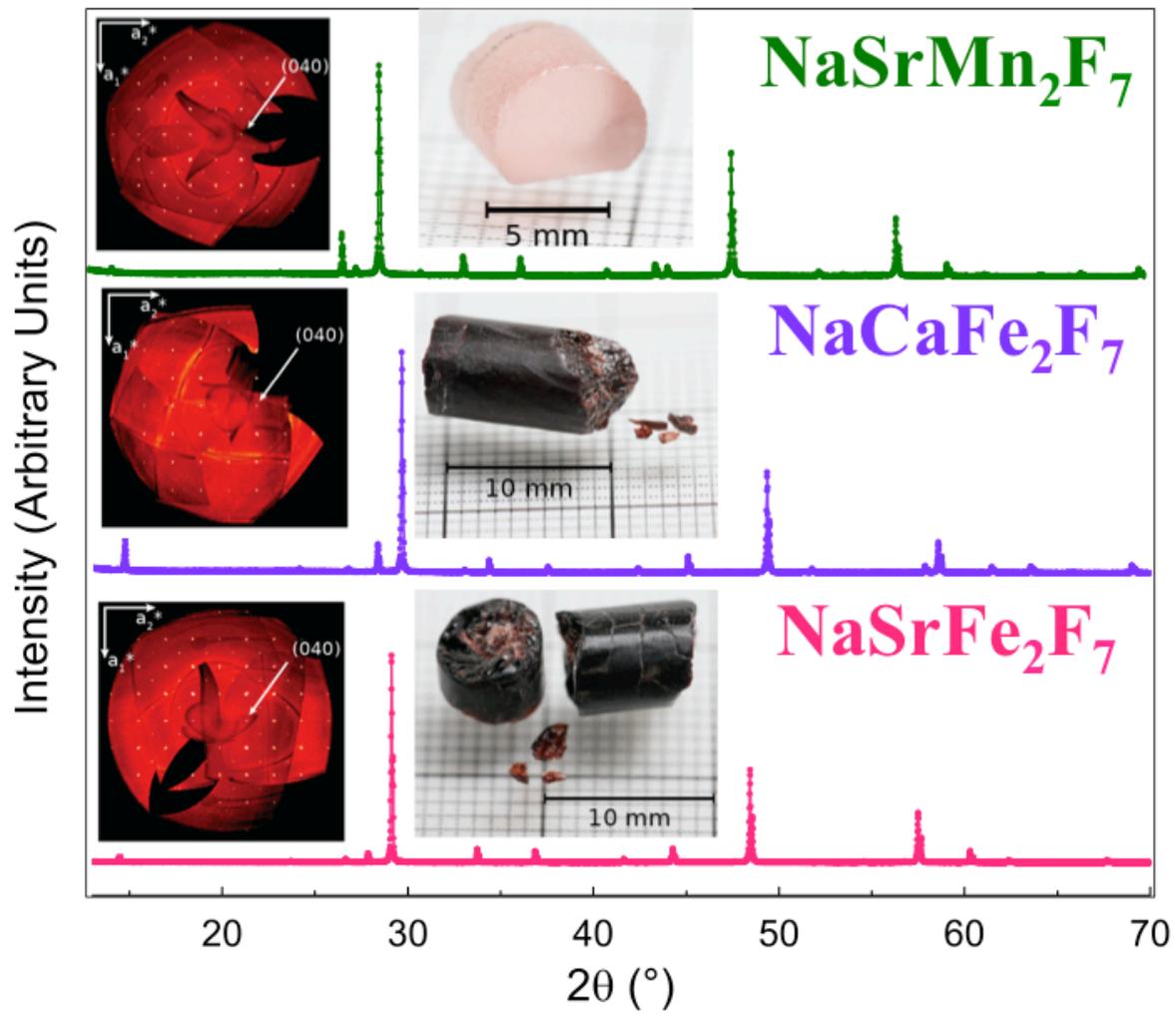

**Figure 2**

# $NaSrFe_2F_7$

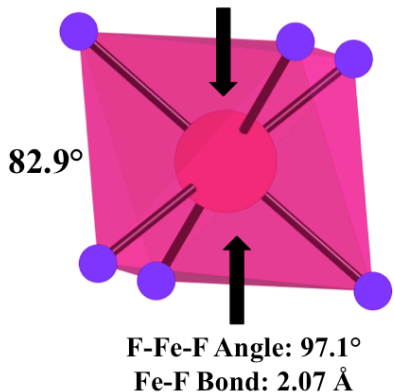

**Octahedra flattened on <111> direction**

82.9°

**F-Fe-F Angle: 97.1°**
**Fe-F Bond: 2.07 Å**

**Figure 3**

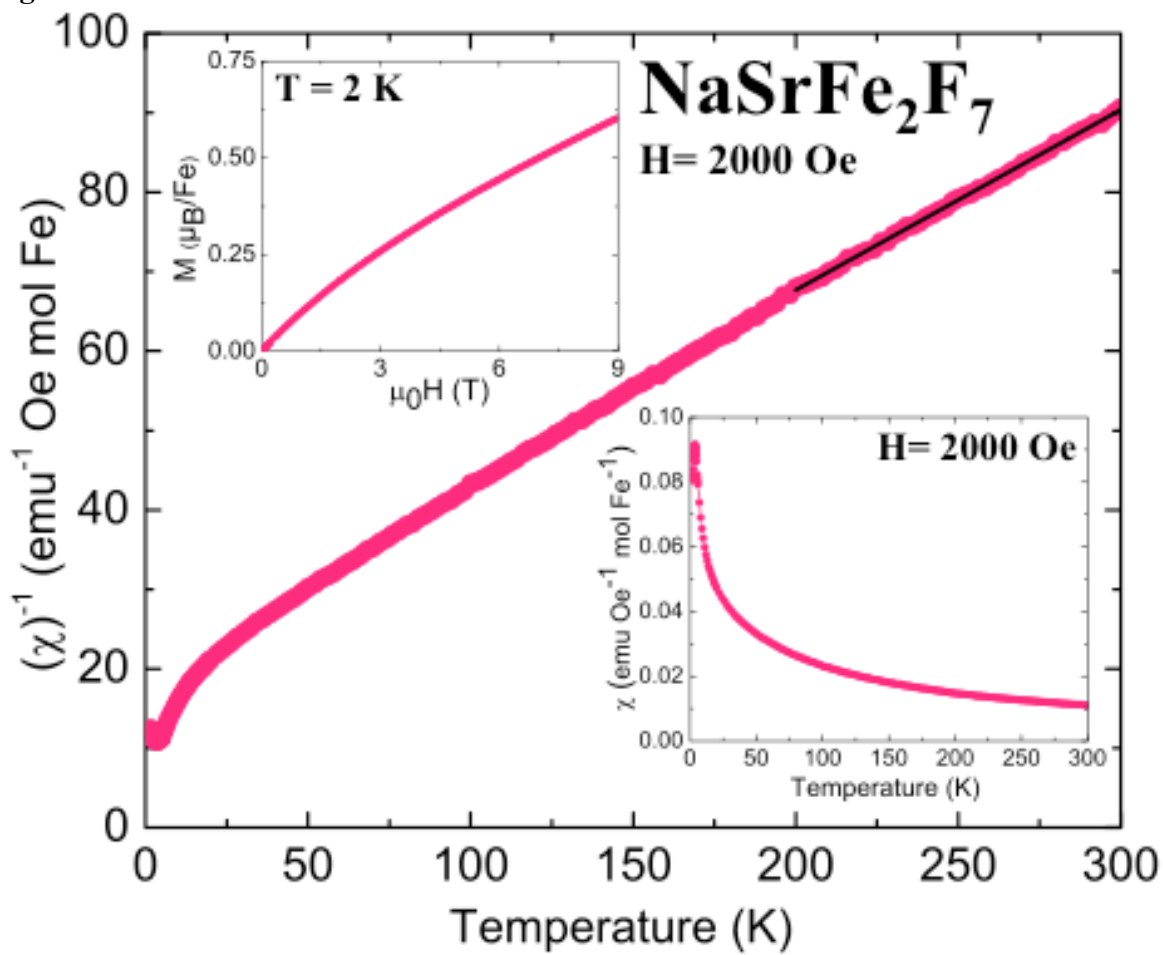

**Figure 4**

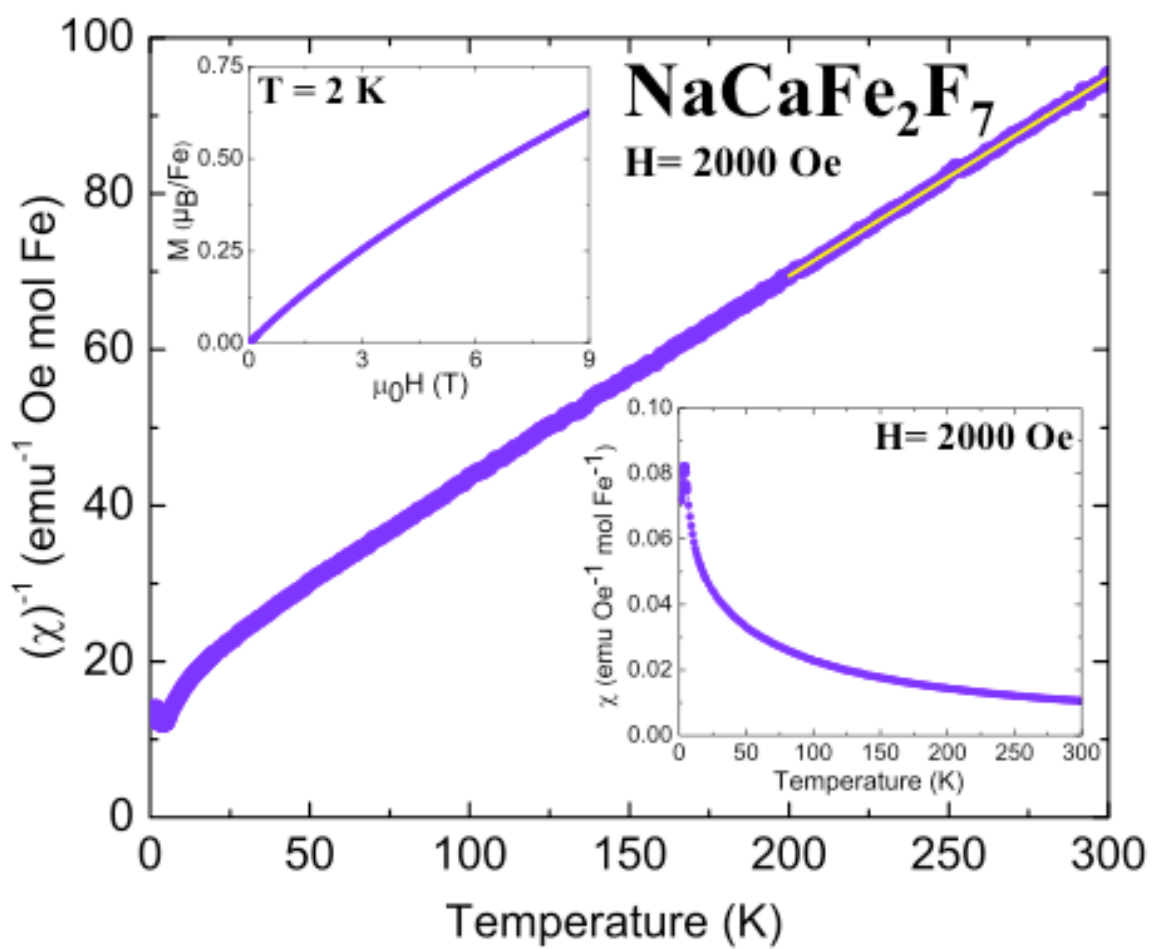

**Figure 5**

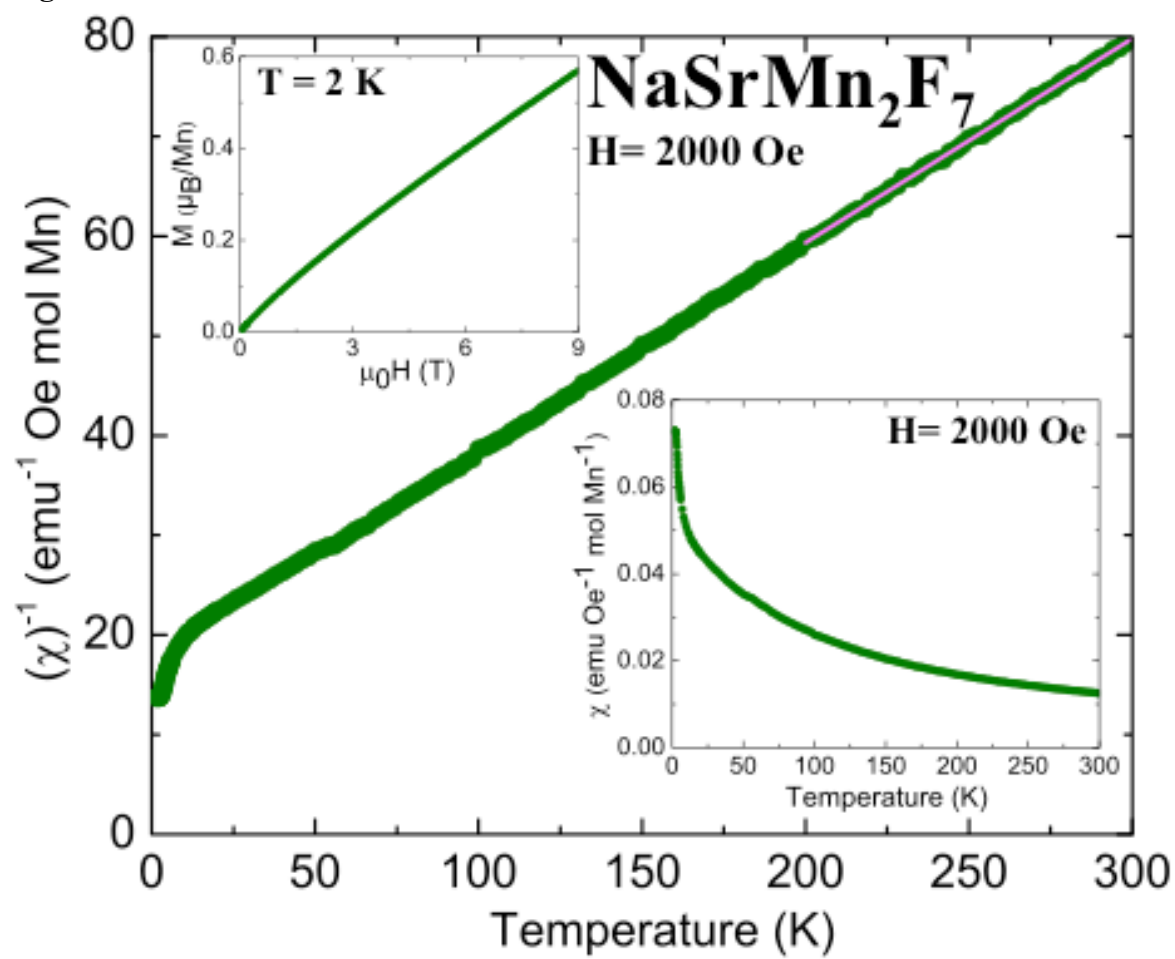

**Figure 6**

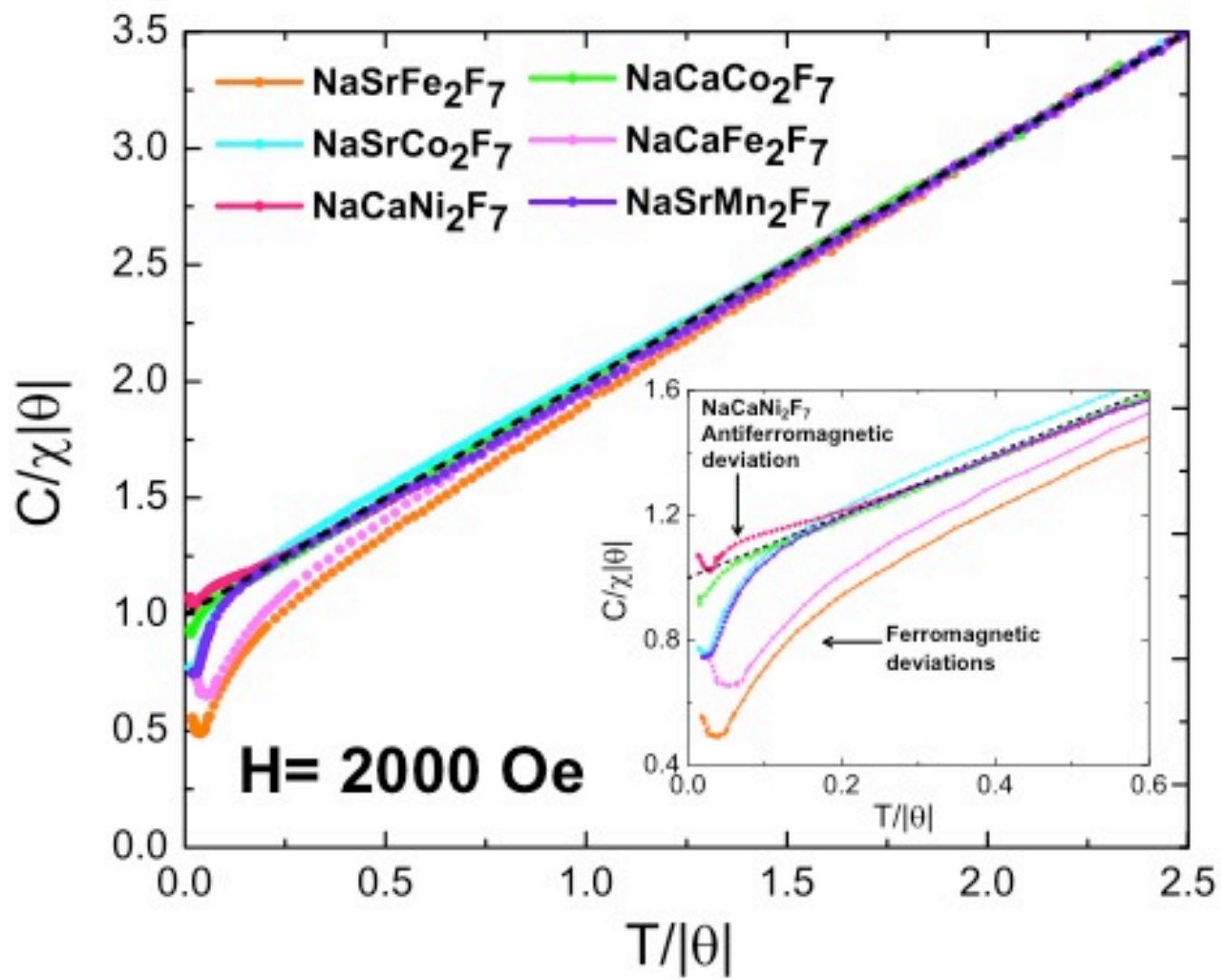

**Figure 7**

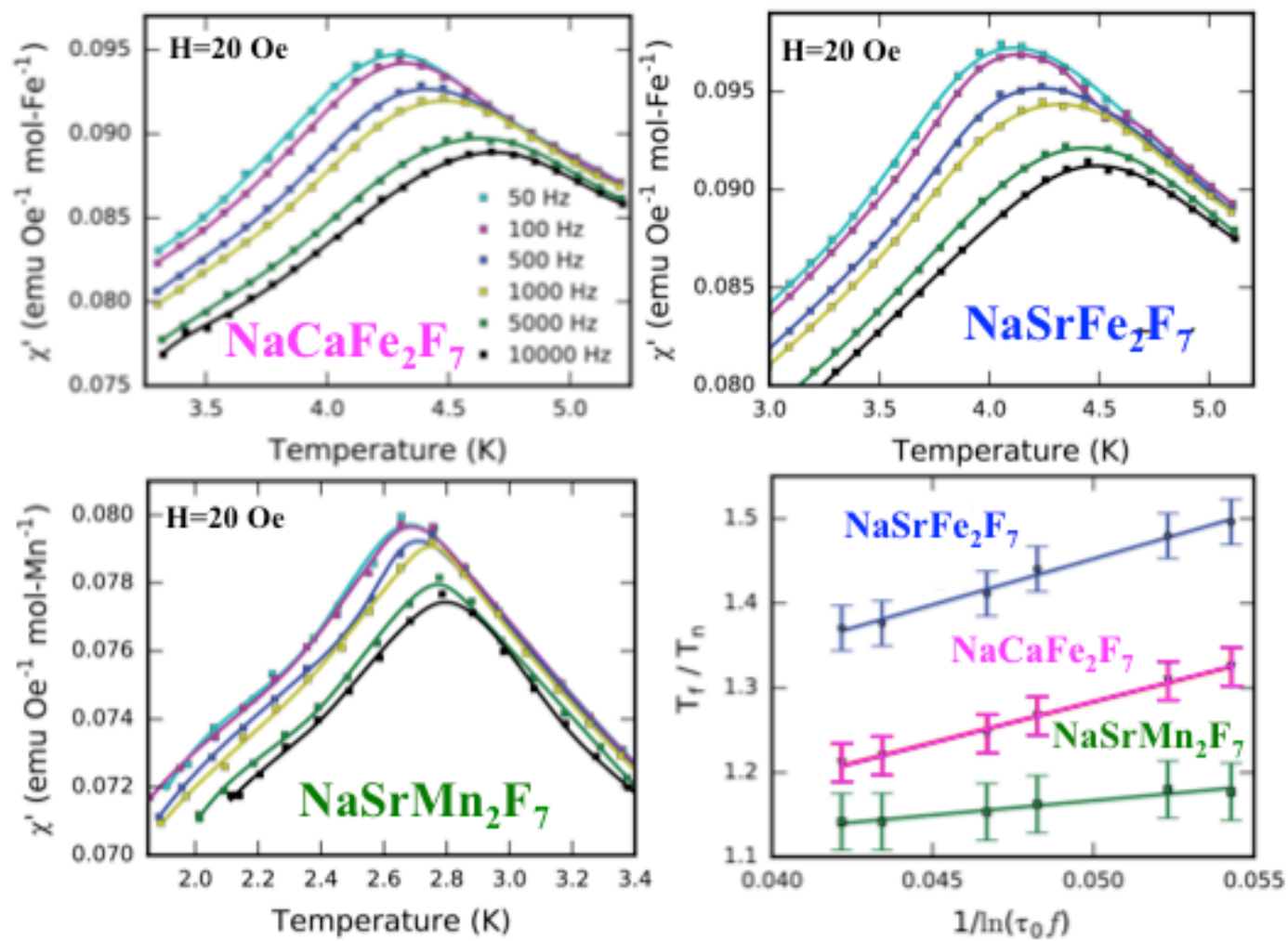

**Figure 8**

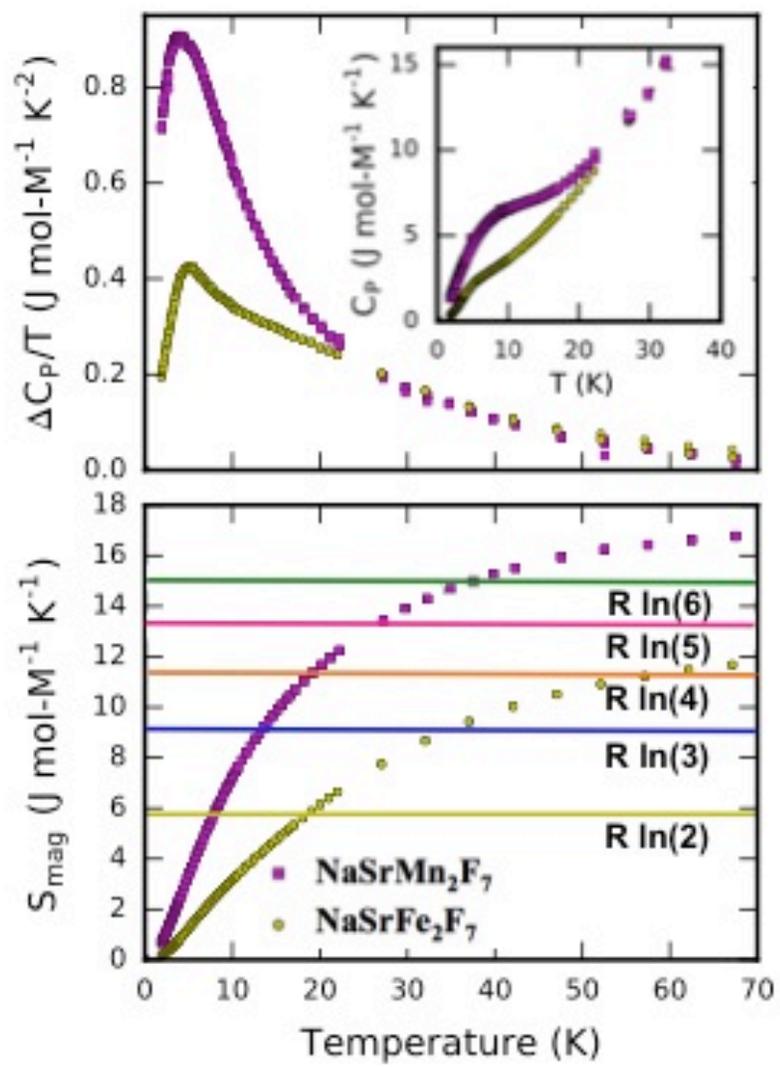